\documentclass[aps,prd,twocolumn,showkeys,superscriptaddress]{revtex4-2}

\usepackage{graphicx}
\usepackage{amsmath}
\usepackage{epstopdf}
\usepackage{amsfonts,amssymb}
\usepackage[utf8]{inputenc}
\usepackage{pstricks}
\usepackage{color}
\usepackage{multirow}
\usepackage{booktabs}
\usepackage{lipsum}

\usepackage[compat=1.0.0]{tikz-feynman}
\usepackage{simplewick}
\usepackage{color, colortbl}
\definecolor{Gray}{gray}{0.9}
\usepackage{float}
\usepackage{tikz-feynman}
\usepackage{feynmp}
\usepackage{forest}

\usepackage{verbatim}    
\usepackage{dcolumn}
\usepackage{bm}
\usepackage{epsfig}
\usepackage{braket}
\usepackage[normalem]{ulem} 
\usepackage{longtable} 

\usepackage{hyperref}

\newcommand{\be}{\begin{equation}}
\newcommand{\ee}{\end{equation}}
\newcommand{\ben}{\begin{eqnarray}}
\newcommand{\een}{\end{eqnarray}}

\newcommand{\pslash}{\not{\hbox{\kern-2.3pt $p$}}}
\newcommand{\pdslash}{\not{\hbox{\kern-2pt $\partial$}}}

\begin{document}


\title{Probing the $G(3900)$ State in Heavy-Ion Collisions}

\author{Luciano M. Abreu}
\email{luciano.abreu@ufba.br}
\affiliation{ Instituto de F\'isica, Universidade Federal da Bahia,
Campus Universit\'ario de Ondina, 40170-115, Bahia, Brazil}

\author{Pedro Brand\~ao}
\email{pedro.brandao@ufba.br}
\affiliation{ Instituto de F\'isica, Universidade Federal da Bahia,
Campus Universit\'ario de Ondina, 40170-115, Bahia, Brazil}

\author{Rodrigo O. Magalh\~aes}
\email{rodrigomagalhaes@ufba.br}
\affiliation{ Instituto de F\'isica, Universidade Federal da Bahia,
Campus Universit\'ario de Ondina, 40170-115, Bahia, Brazil}

\author{Fernando S. Navarra}
\email{navarra@if.usp.br}
\affiliation{Instituto de F\'{\i}sica, Universidade de São Paulo,  
Rua do Mat\~ao 1371,  05508-090
Cidade Universit\'aria, São Paulo, SP, Brazil} 



\begin{abstract}

We investigate the interactions of the recently confirmed exotic hadron $G(3900)$ — with quantum numbers $J^{PC}=1^{--}$ — with the hadronic medium composed of light mesons produced in heavy-ion collisions. Using an effective Lagrangian approach, we compute both vacuum and thermally averaged cross sections for processes such as $G(3900) + \pi \to D^{(\ast)} + \bar{D}^{(\ast)}$ and their inverse reactions. Following recent proposals, the $G(3900)$ is interpreted as a $P$-wave $D\bar{D}^*/D^*\bar{D}$ molecular state. The resulting thermally averaged cross sections are employed in a rate equation to determine the time evolution of the $G(3900)$ multiplicity, with initial conditions provided by statistical and coalescence hadronization models. We compute the number  of $G(3900)$'s 
produced in central Pb-Pb collisions at $\sqrt{s_{NN}} = 5.02$ TeV and 
compare it with the number of produced $Z_c(3900)$'s . We obtain a considerably smaller final yield of $G(3900)$ compared to $Z_c(3900)$.

\end{abstract}

\maketitle

\section{Introduction}
\label{Introduction} 

Since the experimental discovery of the exotic state $X(3872)$ in 2003, numerous hadrons that do not fit comfortably within the original quark model---i.e., the conventional picture of hadrons as $q\bar{q}$ mesons and $qqq$ baryons---have been observed~\cite{brambilla2019xyz,wang2022radiative}. Among these exotic candidates is the vector state $G(3900)$, the focus of the present work.

A peak structure around $3.9$ GeV was first reported nearly two decades ago by the BaBar collaboration in the reaction $e^{+}e^{-} \to D\bar{D}$~\cite{BaBar:2006qlj}, and a similar structure was subsequently observed by the Belle collaboration~\cite{Belle:2007qxm}. More recently, the BESIII collaboration performed high-precision measurements at the BEPCII collider, extracting the Born cross sections for $e^+e^- \rightarrow D^0\bar{D}^0$ and $e^+e^- \rightarrow D^+D^-$. In the resulting energy distribution, a clear structure associated with the $G(3900)$ was identified. Fitting this structure with a Breit–Wigner formalism yields a mass of $m = 3872.5 \pm 14.2 \pm 3.0$ MeV and a width of $\Gamma = 179.7 \pm 14.1 \pm 7.0$ MeV~\cite{BESIII:2024ths}. This precise determination has renewed the theoretical interest in the $G(3900)$.

Since its early observation, various theoretical studies have been carried out to understand the underlying structure of the $G(3900)$. Its possible hadronic molecular nature has been investigated~\cite{Entem:2011nv,Ortega:2012rs,Du:2016qcr,Lin:2024qcq, Chen:2025gxe, Ye:2025ywy, Chen:2025mgg}. In particular, several recent studies based on distinct approaches advocate that this structure corresponds to a $P$-wave molecular state predominantly composed of $D\bar{D}^*/D^*\bar{D}$ mesons~\cite{Lin:2024qcq, Chen:2025gxe, Ye:2025ywy, Chen:2025mgg}. In Ref. \cite{Chen:2025gxe} the authors investigated $S$- and $P$-wave interactions in the $D\bar{D}^*$ system using the Bethe--Salpeter equation, finding that $G(3900)$ is consistent with a $P$-wave molecular state near the $D\bar{D}^*$ threshold, with $I^G(J^{PC}) = 0^-(1^{--})$. In Ref.~\cite{Ye:2025ywy}, a global analysis of the cross sections for $e^+e^- \rightarrow D\bar{D}$, $e^+e^- \rightarrow D\bar{D}^*+c.c.$, and $e^+e^- \rightarrow D^*\bar{D}^*$ suggested that the $G(3900)$ is a $P$-wave dynamically generated state. In ~\cite{Chen:2025mgg} the authors constructed an effective theory for heavy-meson interactions, adjusted the low-energy constants for the $P$-wave channel, and obtained a $P$-wave $D\bar{D}^*/D^*\bar{D}$ resonant state compatible with the $G(3900)$. 

Nevertheless, alternative interpretations for the $G(3900)$ have also emerged. For instance, in Ref.~\cite{Cao:2014qna} it was suggested that this state might not be a genuine resonance but rather a cusp effect originating from the opening of the $D^*\bar{D} + c.c.$ channel. Using a coupled-channel formalism that incorporates both bare $c\bar{c}$ and hadronic channels, the authors of Ref.~\cite{Qian:2025zyp} interpreted the $G(3900)$ as a structure arising from such channel dynamics. Meanwhile, Ref.~\cite{Cao:2025okk} simulated $G(3900)$ production in $e^+e^-$ collisions using the PACIAE parton/hadron cascade model combined with a dynamically constrained phase-space coalescence model, assuming either a $P$-wave $D\bar{D}$ or $D\bar{D}^{*}/\bar{D}D^{*}$ molecular interpretation. Their results indicate that the $G(3900)$ multiplicity in the $D\bar{D}^{*}/\bar{D}D^{*}$ scenario is significantly larger than in the $D\bar{D}$ case. From yet other perspectives, Refs.~\cite{Husken:2024hmi,Salnikov:2024wah,Huang:2025g3900,Huang:2025xvv} propose that the $G(3900)$ could be a kinematic effect or an excited $2p$ state. In particular, Ref.~\cite{Huang:2025g3900} suggests the production of $G$ via a triangle singularity mechanism, where it is suggested that $G$ could arise from a $D_1 \bar D D^*$ triangle loop, similarly to the $Z_c(3900)$.

In view of the ongoing debate about the $G(3900)$ nature, additional investigation is clearly needed to better determine its properties and internal composition. In this context, heavy-ion collisions (HICs) provide a particularly useful testing ground. In the initial stages of these collisions a large
number of heavy quarks are produced. As the system evolves  in the quark-gluon plasma (QGP) phase, it expands and cools, transitioning into a hot hadron gas at hadronization. During this process, heavy quarks recombine into bound states, which include not only ordinary mesons and baryons but also multiquark configurations. This expectation gained strong support from the CMS observation of the $X(3872)$ in Pb-Pb collisions at $\sqrt{s_{NN}} = 5.02$ TeV, which revealed a prompt production rate roughly an order of magnitude higher than in $p$-$p$ collisions~\cite{CMS:2021znk}. Moreover, the LHCb Collaboration has reported the $X(3872)$ in $p$-Pb collisions at $\sqrt{s}=8.16$ TeV per nucleon~\cite{LHCb:2024bpb}.

On the theoretical front, numerous investigations have recently emerged~\cite{Esposito:2020ywk,Wu:2020zbx,Zhang:2020dwn,Yun:2022evm}. Our group has dedicated a 
continuous effort to study exotic hadrons in HICs. Earlier estimates of both conventional and selected exotic hadron yields in these collisions relied on statistical and coalescence models that omitted interactions occurring in the hadron gas phase~\cite{EXHIC}. However, our subsequent studies have demonstrated that hadronic interactions can substantially alter the final abundances (see, e.g., Refs.~\cite{XProd1,ABREU2016303,Abreu:2017cof,Abreu:2018mnc,Abreu:2020ony,chiara,abreu2022exotic,abreu2022interactions,Abreu:2022lfy,Abreu1,Abreu:2023jcs,Abreu:2023awj,Abreu:2024mxc}).  Heavy hadrons produced during hadronization may be suppressed through scattering with other medium constituents, or alternatively, they may be produced via the recombination of lighter particles. These medium interactions are especially interesting in the case of exotic states, as molecular and compact tetraquark configurations exhibit different spatial sizes, leading to different interaction cross sections. Therefore, in principle, the medium could act 
as a "filter" of a given configuration. 

In the present work, we take a further step within the framework of our ongoing program. We examine the interactions of the $G(3900)$ with light mesons present in the hadronic medium formed in heavy-ion collisions. To this end, we employ an effective Lagrangian formalism to compute cross sections in vacuum as well as thermally averaged ones for the reactions $G(3900)\,\pi \to D^{(*)}\bar{D}^{(*)}$ and their inverse processes, assuming that the $G(3900)$ is a $P$-wave $D\bar{D}^*/D^*\bar{D}$ molecular state. These thermally averaged cross sections are then fed into a rate equation that tracks the $G(3900)$ multiplicity over time, with initial conditions taken from statistical and coalescence hadronization models. 
Then we compare  $G(3900)$ production with $Z_c(3900)$ production. This comparison is interesting because the two states have the same mass and the 
same quark content, differing only because the former is in a P wave while the latter is in a S wave.

This work is structured as follows. Sec.~\ref{amplitude} describes the calculation and reports the vacuum and thermally averaged cross sections. Sec.~\ref{Abundance} presents the time evolution of the $G(3900)$ multiplicity. Finally, Section~\ref{Conclusions} is devoted to the discussion and concluding remarks. In the Appendix, we detail the calculations leading to the coupling constant $g_{G_0DD^{*}}$.

\section{$G(3900)$ interactions}
\label{amplitude}

\subsection{Transition amplitudes}

\begin{figure} [!ht]
\begin{align*}
\begin{tikzpicture}
\begin{feynman}
\vertex (a1) {$G (p_1)$};
	\vertex[right=1.5cm of a1] (a2);
	\vertex[right=1.cm of a2] (a3) {$\bar{D} (p_3)$};
	\vertex[right=1.4cm of a3] (a4) {$G (p_1)$};
	\vertex[right=1.5cm of a4] (a5);
	\vertex[right=1.cm of a5] (a6) {$\bar{D} (p_{3})$};
\vertex[below=1.5cm of a1] (c1) {$\pi (p_2)$};
\vertex[below=1.5cm of a2] (c2);
\vertex[below=1.5cm of a3] (c3) {$D (p_4)$};
\vertex[below=1.5cm of a4] (c4) {$\pi (p_2)$};
\vertex[below=1.5cm of a5] (c5);
\vertex[below=1.5cm of a6] (c6) {$D (p_4)$};
	\vertex[below=2cm of a2] (d2) {(a)};
	\vertex[below=2cm of a5] (d5) {(b)};
\diagram* {
(a1) -- (a2), (a2) -- (a3), (c1) -- (c2), (c2) -- (c3), (a2) -- [fermion, edge label'= $D^{*}$] (c2), (a4) -- (a5), (a5) -- (c6), (c4) -- (c5), (c5) -- (a6), (a5) -- [fermion, edge label'= $\bar{D}^{*}$] (c5)
}; 
\end{feynman}
\end{tikzpicture}
\end{align*}

\begin{align*}
\begin{tikzpicture}
\begin{feynman}
\vertex (a1) {$G (p_1)$};
	\vertex[right=1.5cm of a1] (a2);
	\vertex[right=1.cm of a2] (a3) {$\bar{D}^{*} (p_3)$};
	\vertex[right=1.4cm of a3] (a4) {$G (p_1)$};
	\vertex[right=1.5cm of a4] (a5);
	\vertex[right=1.cm of a5] (a6) {$\bar{D}^{*} (p_{3})$};
\vertex[below=1.5cm of a1] (c1) {$\pi (p_2)$};
\vertex[below=1.5cm of a2] (c2);
\vertex[below=1.5cm of a3] (c3) {$D^{*} (p_4)$};
\vertex[below=1.5cm of a4] (c4) {$\pi(p_2)$};
\vertex[below=1.5cm of a5] (c5);
\vertex[below=1.5cm of a6] (c6) {$D^{*} (p_4)$};
	\vertex[below=2cm of a2] (d2) {(c)};
	\vertex[below=2cm of a5] (d5) {(d)};
\diagram* {
(a1) -- (a2), (a2) -- (a3), (c1) -- (c2), (c2) -- (c3), (a2) -- [fermion, edge label'= $D$] (c2), (a4) -- (a5), (a5) -- (c6), (c4) -- (c5), (c5) -- (a6), (a5) -- [fermion, edge label'= $\bar{D}$] (c5)
}; 
\end{feynman}
\end{tikzpicture}
\end{align*}

\begin{align*}
\begin{tikzpicture}
\begin{feynman}
\vertex (a1) {$G (p_1)$};
	\vertex[right=1.5cm of a1] (a2);
	\vertex[right=1.cm of a2] (a3) {$D (p_3)$};
\vertex[below=1.5cm of a1] (c1) {$\pi (p_2)$};
\vertex[below=1.5cm of a2] (c2);
\vertex[below=1.5cm of a3] (c3) {$\bar{D}^{*} (p_4)$};
	\vertex[below=2cm of a2] (d2) {(e)};
\diagram* {
(a1) -- (a2), (a2) -- (a3), (c1) -- (c2), (c2) -- (c3), (a2) -- [fermion, edge label'= $\bar{D}^{*}$] (c2)
}; 
\end{feynman}
\end{tikzpicture}
\end{align*}
\caption{Diagrams contributing to the following processes: $G\pi \to D\bar{D}$ (\textbf{a}) and (\textbf{b}), $G\pi \to D^*\bar{D}^*$ (\textbf{c}) and (\textbf{d}), and $G\pi \to D\bar{D}^*$ (\textbf{e}). The charges of the particles are not specified. The momenta of the particles in the  initial (final) state are denoted as $p_1$ and $p_2$ ($p_3$ and $p_4$).}
\label{fig:diagrams}
\end{figure}
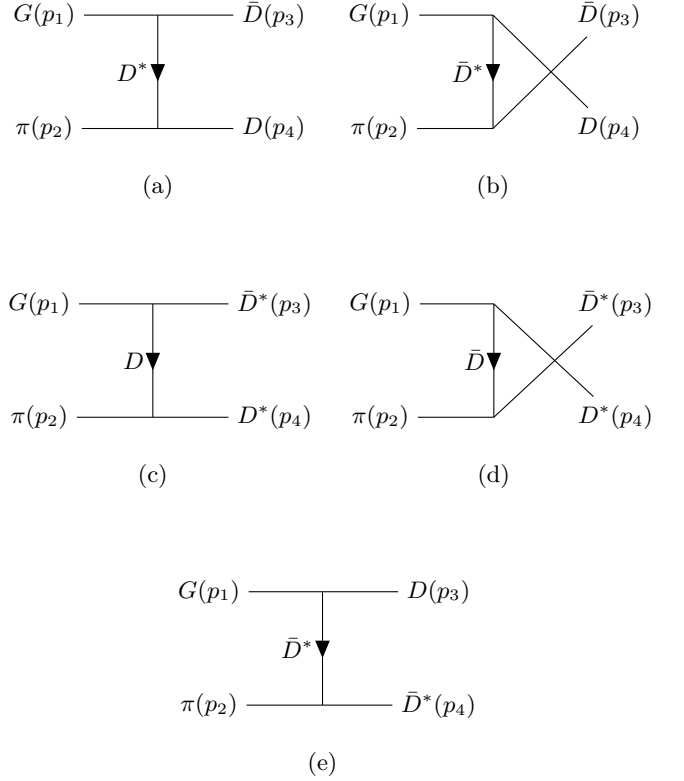

We describe the interactions of the $G(3900)$ state (henceforth denoted as $G$) with a pion-dominated hadronic medium using the effective Lagrangian formalism. More precisely, the $G$ suppression via pion-induced reactions is modeled at the lowest-order Born level for the processes $G \pi \to D\bar{D}$, $G \pi \to D^*\bar{D}^*$, and $G \pi \to D\bar{D}^*$, as illustrated in Fig.~\ref{fig:diagrams}. The effective Lagrangians governing the relevant vertices are given by~\cite{cho2013hadronic,abreu2022interactions,Huang:2025xvv}:
\begin{align}
\label{lagrangianas}
&\mathcal{L}_{GDD^*} = -g_{GDD^*}\epsilon^{\mu\nu\rho\sigma} 
\partial_\mu G_\nu
D \overset{\leftrightarrow}{\partial_\rho} \bar{D}^*_\sigma + h.c., \nonumber \\
	&\mathcal{L}_{\pi DD^*} = ig_{\pi D D^*} D^{*}_\mu \vec{\tau} \cdot (\bar{D} \partial^\mu \vec{\pi} - \partial^{\mu} \bar{D} \vec{\pi}) + h.c.,  \nonumber \\
	&\mathcal{L}_{\pi D^*D^*}=-g_{\pi D^*D^*}\varepsilon^{\mu\nu\alpha\beta}\partial_\mu D^*_\nu \pi \partial_\alpha\bar{D}^*_\beta,
\end{align}
where $\vec{\pi}$ represents the pion isospin triplet; $\vec{\tau}$ denotes the Pauli matrices in isospin space; $D^{(*)}$ and $\bar D^{(*)}$ are the isospin doublets for the pseudoscalar (vector) charmed mesons; $G_\nu$ represents the vector state $G(3900)$; and $g_{GDD^*}$, $g_{\pi D D^*}$, and $g_{\pi D^* D^*}$ are coupling constants.  The first two are given in Refs.~\cite{Abreu:2024mxc,magalhaes_2024}:  $g_{\pi DD^*} = 6.3$ and $g_{\pi D^*D^*} = 9.08$ GeV$^{-1}$. The estimation of $g_{GDD^*}$ encodes the interpretation of the $G(3900)$ as a $P$-wave $D\bar{D}^{*}/\bar{D}D^{*}$ molecular state~\cite{Ye:2025ywy}. Accordingly, $g_{GDD^*}$ is determined via the compositeness condition~\cite{Weinberg:1965zz}. The details of this calculation are presented in Appendix~\ref{appa}.

With these effective Lagrangians,  the scattering amplitudes associated to the reactions shown in Fig.~\ref{fig:diagrams} are given by
\begin{align} \label{eq: amplitude sum}
&\mathcal{M}(G \pi \to D\bar{D}) = \mathcal{M}^{(a)}+\mathcal{M}^{(b)}, \nonumber \\
&\mathcal{M}(G \pi \to D^*\bar{D}^*) = \mathcal{M}^{(c)}+\mathcal{M}^{(d)}, \nonumber \\
&\mathcal{M}(G \pi \to D\bar{D}^*) = \mathcal{M}^{(e)},
\end{align}
where ${M}^{(i)}$ represents the amplitude coming from the specific process $i=a,..,e$; these expressions are explicitly written as 
\begin{eqnarray}
\mathcal{M}^{(a)}& = & -g_{GDD^*} g_{\pi DD^*}\; \varepsilon^{\mu\nu\rho\sigma}\,p_{1\mu}\,\varepsilon_\nu(p_1,\lambda)\; \nonumber \\ & & \times \left(g_{\alpha\sigma}-\dfrac{(p_1-p_3)_\alpha (p_1-p_3)_\sigma}{m_{D^*}^2}\right) \nonumber
 \\ & &  
\times \frac{1}{t-m_{D^*}^2}\;(p_1-2p_3)_\rho\,(p_2+p_4)^\alpha , \nonumber \\ 
\mathcal{M}^{(b)} & = &  -g_{GDD^*} g_{\pi DD^*}\; \varepsilon^{\mu\nu\rho\sigma}\,p_{1\mu}\,\varepsilon_\nu(p_1,\lambda)\; \nonumber \\ && \times \left(g_{\alpha\sigma}-\dfrac{(p_1-p_4)_\alpha (p_1-p_4)_\sigma}{m_{D^*}^2}\right)
		\nonumber \\ && \times \frac{1}{u-m_{D^*}^2}\;
		(p_1-2p_4)_{\rho}\,(p_2+p_3)^\alpha  , \nonumber \\ 
\mathcal{M}^{(c)} & =& g_{GDD^*}\,g_{\pi D D^*}\;
		\varepsilon^{\mu\nu\rho\sigma}\,p_{1\mu}\,\varepsilon_\nu(p_1,\lambda)\;
		\left(\frac{1}{t-m_D^2}\right) \nonumber \\ &&\times\;(p_1-2p_3)_\rho\,(p_4-2p_2)^\alpha\;
		\varepsilon^*_\sigma(p_3,\lambda')\,\varepsilon^*_\alpha(p_4,\lambda'') , \nonumber \\
\mathcal{M}^{(d)} & = & \,g_{GDD^*}\,g_{\pi D D^*}\,
	\varepsilon^{\mu\nu\rho\sigma}\,p_{1\mu}\,\varepsilon_\nu(p_1,\lambda)\,
	\left(\frac{1}{u-m_D^2}\right)\; \nonumber \\ && \times (p_1-2p_4)_\rho\,(p_3-2p_2)^\alpha\,
	\varepsilon^*_\sigma(p_4,\lambda'')\,\varepsilon^*_\alpha(p_3,\lambda') , \nonumber \\
\mathcal{M}^{(e)} & = &  g_{GDD^*} g_{\pi D^*D^*} \epsilon^{\mu\nu\rho\sigma}\epsilon^{\gamma\theta\delta\eta}\varepsilon_\nu(p_1,\lambda)\varepsilon_\theta(p_4,\lambda')  \nonumber \\ && \times \left( g_{\sigma \eta}-\dfrac{(p_1-p_3)_{\sigma} (p_1-p_3)_{\eta}}{m_{\bar{D}^{*}}^2} \right) \left(\frac{1}{t-m_{\bar{D}^{*}}^2}\right) \nonumber \\ &&\times
p_{1\mu}p_{4\gamma}(2p_3-p_1)_\rho(p_4-p_2)_\delta,
\end{eqnarray}
where $s, u$ and $t$ are the Mandelstam variables, defined as $ s=(p_1+p_2)^2, t=(p_1-p_3)^2 $ and $ u=(p_1-p_4)^2$; and  $\epsilon^{(\ast)} (p_i,\lambda^a)$ is the polarization vector associated to the corresponding vector meson.


\subsection{Vacuum Cross Sections}
\label{CrossSection}

The isospin- and spin-averaged cross section for the reactions in Eq.~(\ref{eq: amplitude sum}) in the center-of-mass (CM) frame is given by
\begin{equation} \label{seção de choque total}
\sigma_{ab \to cd} = \frac{1}{64 g_a g_b \pi^2 s} \frac{|\vec{p}_{cd}|}{|\vec{p}_{ab}|} \int d\Omega \, \overline{\sum} |\mathcal{M}_{ab \to cd} |^2,
\end{equation}
where $g_{a,b} = (2I_{a,b}+1)(2S_{a,b}+1)$ is the degeneracy factor of the initial-state particles; $s$ is the squared CM energy; $|\vec{p}_{ab}|$ and $|\vec{p}_{cd}|$ are the absolute values of the three-momenta of the initial and final particles in the CM frame; and $\overline{\sum}$ denotes summation over the spin and isospin of the initial and final states. This summation can be rewritten in the particle basis as
\begin{equation}
 \displaystyle  \sum_{I} |M_{ab\rightarrow cd}|^2 \rightarrow \sum_{Q_c,Q_d} |M_{ab\rightarrow cd}^{(Q_c,Q_d)}|^2,
\end{equation}
where $Q_c$ and $Q_d$ are the explicit charges of the particles in the final state.

The inverse processes involving the production of the $G(3900)$, namely $D\bar{D} \to G\pi$, $D^*\bar{D}^* \to G\pi$, and $D\bar{D}^* \to G\pi$, the cross sections are obtained using the detailed balance relation:
\begin{equation}
g_a g_b |\vec{p}_{ab}|^2 \sigma_{ab \to cd}(s) = g_c g_d |\vec{p}_{cd}|^2 \sigma_{cd \to ab}(s).
\label{detbal}
\end{equation}

To take into account the finiteness of hadrons and also prevent the artificial growth of the cross sections with the energy, we employ a monopole-type form factor~\cite{tcc22},
\begin{equation}
    F(\mathbf{q})=\frac{\Lambda^2}{\Lambda^2 + \mathbf{q}^2},
\end{equation}
where $\mathbf{q}$ denotes the three-momentum transferred by the virtual particles in the $t$ and $u$ channels, and the cutoff parameter $\Lambda$ is taken to be $2.0~\text{GeV}$. 

We use the isospin-averaged values for the masses the mesons reported in~\cite{ParticleDataGroup:2024cfk}; for the $G(3900)$ we use $m_{G} = 3.8725 \,\text{GeV}$~\cite{BESIII:2024ths}. The results are shown as bands that reflect the range between the smallest and largest allowed values of the coupling constant $g_{GDD^*}$ discussed in Appendix~\ref{appa}.

\begin{figure}[htb!]
\centering
\includegraphics[{width=0.9\linewidth}]{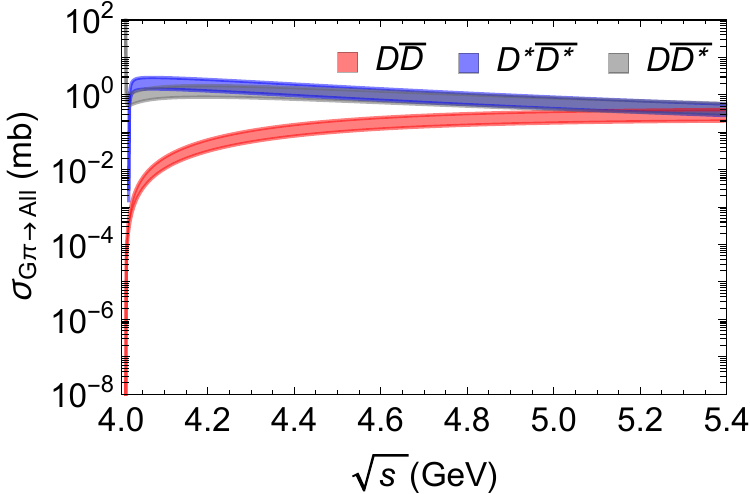}\\ 
\vskip0.5cm
\includegraphics[{width=0.9\linewidth}]{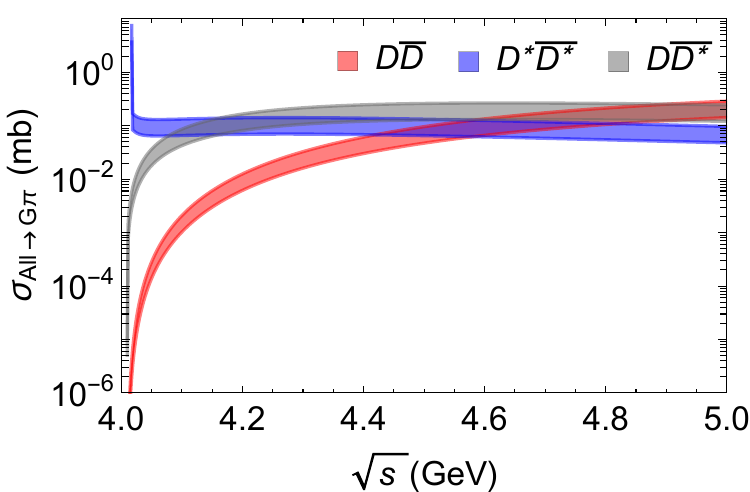}
\caption{Top: cross sections for the suppression processes $G \pi \to D^{(*)}\bar{D}^{(*)}$ as functions of $\sqrt{s}$. Bottom: cross sections for the corresponding inverse reactions. } 
    \label{Fig: CrossSection}
\end{figure}

Fig.~\ref{Fig: CrossSection} presents the cross sections for the $G(3900)$ suppression processes as a function of $\sqrt{s}$. Near the threshold,  the cross section for the process process $G\pi \to D\bar{D}$ is suppressed with respect to the other reactions. The most relevant contribution at moderate CM energies is given by the reaction $G\pi \to D^*\bar{D}^*$. 

\begin{figure}[h!]
\centering
\includegraphics[{width=0.9\linewidth}]{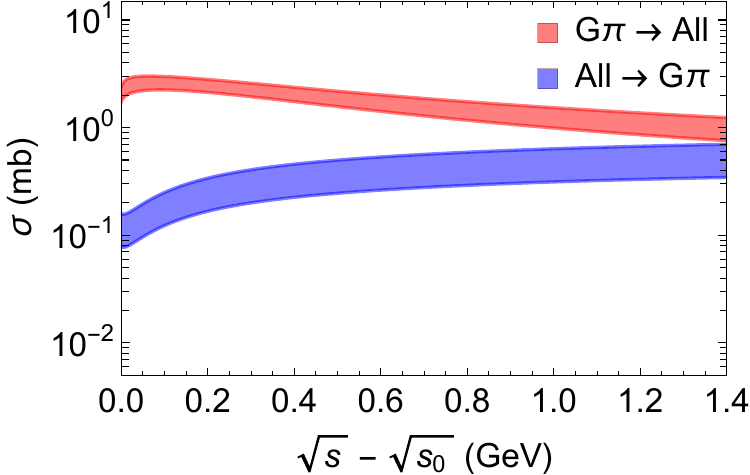}
\caption{Sum of all cross-sections for the $G$ absorption and production processes as functions of $\sqrt{s}-\sqrt{s_0}$, 
where $\sqrt{s_0}$ is the corresponding threshold of each channel. $All $ represents the sum 
$\sigma(G \pi \to D\bar{D})+ \sigma(G \pi \to D^*\bar{D}^*)+\sigma(G \pi \to D\bar {D}^*)$ for the case of absorption processes 
and an equivalent  expression for the inverse ones. }
\label{fig: CrossSectionSum}
\end{figure}

Fig.~\ref{Fig: CrossSection} also shows the cross sections for the inverse reactions, i.e., the production processes. The reactions with initial states $D\bar{D}$ and $D\bar{D}^*$ are endothermic, while the process with initial state $D^{*}\bar{D}^*$ is exothermic, as expected. Again, the cross section for the process $D\bar{D} \to G\pi$ in the region close to the threshold is the least relevant contribution. On the other hand, the processes with initial state $D\bar{D}^*$ and $D^*\bar{D}^*$ have cross sections of the same order of magnitude.
 
Comparing the absorption and production processes of $G$ in the energy region relevant for heavy ion collisions ($\sqrt{s} - \sqrt{s_0} < 0.6$~GeV), the cross section of the $G\pi \to D\bar{D}$ process in this region is approximately of the same order of magnitude as its inverse process, near ($\sqrt{s} - \sqrt{s_0} = 0.6$~GeV). For values lower than 0.6, the absorption process is greater than the production process. The other absorption processes are about one order of magnitude larger than their inverse processes, with this difference being slightly greater for processes involving the initial and final $D^*\bar{D}^*$ states.

To better compare these contributions, Fig.~\ref{fig: CrossSectionSum} shows the sum of all cross sections for $G$ suppression and production as a function of $(\sqrt{s} - \sqrt{s_0})$. In the energy region relevant for heavy-ion collisions, i.e., $\sqrt{s} - \sqrt{s_0} \leq 0.6$ GeV, the total suppression cross section is larger than that of the production processes, differing by approximately one order of magnitude.


\subsection{Thermally-averaged cross sections}
\label{ThermalAverage}

In a heavy-ion collision environment, where medium effects are relevant and the $G(3900)$ can interact with light hadrons, the collision energy is intrinsically related to the temperature of the medium, making the calculation of thermally averaged cross sections essential. This thermal average is defined as the convolution of the vacuum cross section with the momentum distributions, i.e., by~\cite{magalhaes_2024,cho2013hadronic,ABREU2016303,Abreu:2022lfy}:
\begin{align}\label{medias termicas}
\langle \sigma_{ab \to cd} v_{ab} \rangle =& \frac{\int d^3\textbf{p}_a \, d^3\textbf{p}_b \, f_a(\textbf{p}_a) f_b(\textbf{p}_b) \, \sigma_{ab \to cd} \, v_{ab}}{\int d^3\textbf{p}_a \, d^3\textbf{p}_b \, f_a(\textbf{p}_a) f_b(\textbf{p}_b)} \nonumber \\
=& \frac{1}{4 \alpha_a^2 K_2(\alpha_a) \, \alpha_b^2 K_2(\alpha_b)} \int_{z_0}^{\infty} dz \, K_1(z) \, \sigma(z T) \nonumber \\
&\times \left[z^2 - (\alpha_a + \alpha_b)^2\right] \left[z^2 - (\alpha_a - \alpha_b)^2\right],
\end{align}
where $v_{ab}$ is the initial relative velocity of particles $a$ and $b$; $f_i(\textbf{p}_i)$ the momentum distribution; $\alpha_i = m_i/T$; $z_0 = \max(\alpha_a + \alpha_b, \alpha_c + \alpha_d)$; and $K_{1,2}$ are modified Bessel functions.

\begin{figure}[htb!]
\centering
\includegraphics[{width=0.9\linewidth}]{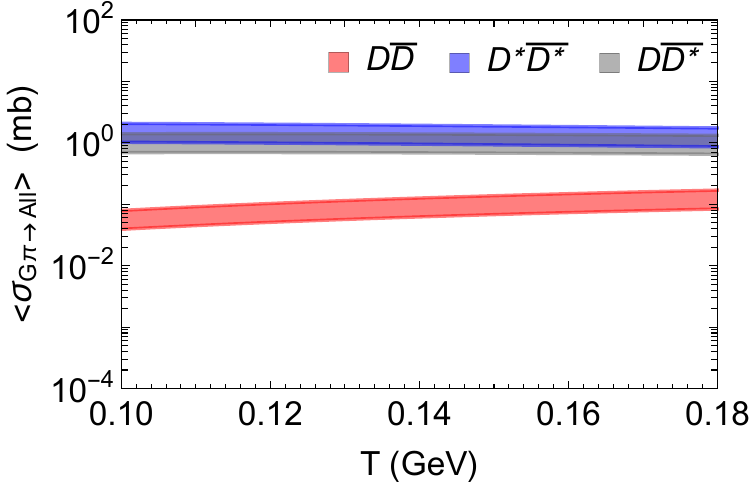}\\
\vskip0.5cm
\includegraphics[{width=0.9\linewidth}]{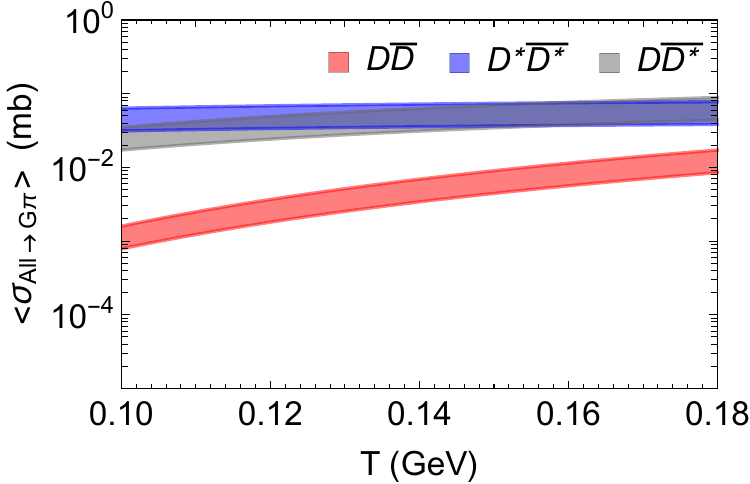}
\caption{Top: Thermally-averaged cross-sections of the processes $G\pi \to D^{(*)}\bar{D}^{(*)}$ as functions of the temperature. 
Bottom:  thermally-averaged cross-sections of the corresponding inverses reactions.}
\label{fig:AvCrsecZcPiAll(Inv)}
\end{figure}

Fig.~\ref{fig:AvCrsecZcPiAll(Inv)} shows the thermally averaged cross sections as functions of temperature for the $G$ suppression and production processes. The suppression processes exhibit only a mild, nearly constant temperature dependence. Among them, the $D\bar{D}$ channel shows the largest (though still small) variation but is the least prominent, with cross sections about two orders of magnitude smaller than the others. The $D^*\bar{D}^*$ channel is the most prominent.
The production processes display a slightly stronger temperature dependence, particularly for the $D\bar{D}$ initial state. 
Importantly, the suppression cross sections are approximately two orders of magnitude larger than the production ones for the dominant $D\bar{D}^*$ and $D^*\bar{D}^*$ channels.

We therefore conclude that absorption processes dominate over production, as observed for most other exotic hadrons~\cite{magalhaes_2024}. This suggests that the hadronic medium might significantly affect the $G$ multiplicity in the late stages of heavy-ion collisions. We will now incorporate these results into a kinetic equation to determine the $G$ yield.

\section{The $G (3900)$ Abundance}
\label{Abundance}

\subsection{Kinetic equation}

Let us now study how the $G(3900)$ multiplicity evolves in time during the hadronic phase of heavy-ion collisions. The previously estimated thermally-averaged cross sections will be input into the following momentum-integrated evolution equation~\cite{Abreu1,XProd1,ABREU2016303,Abreu:2017cof,Abreu:2018mnc,magalhaes_2024}:
\begin{eqnarray} 
	\frac{ d N_{G} (\tau)}{d \tau}  &=& 
	\sum_{\substack{c = D, D^* \\ \bar c = \bar{D}, \bar{D}^{*}}} 
	\left[ \langle \sigma_{c \bar c \rightarrow G \pi } 
	v_{c \bar c } \rangle \, n_{\bar{c}} (\tau) \, N_{c}(\tau)  \right.\nonumber\\ && \left.- \langle \sigma_{ G \pi\rightarrow c \bar c } v_{ G \pi } \right.
	\rangle \, n_{\pi} (\tau) \, N_{G}(\tau) \nonumber \\ 
    & &  \left. - \langle \Gamma_{G} \rangle N_{G} (\tau)  \right],
	\label{rateequation}
\end{eqnarray}
where \(N_{G}(\tau)\) is the $G$ multiplicity of at proper time \(\tau\); \(n_c(\tau)\), \(n_{\bar c}(\tau)\), and \(n_\pi(\tau)\) are the number densities of charmed mesons and pions, respectively; $ \Gamma_{G}$ is the $G$ decay width. 

We assume that the pions and charmed mesons involved in the reactions discussed above are in thermal equilibrium, having their respective densities obeying the Boltzmann distribution,
\begin{align}
	n_{i} (\tau) \approx  \frac{1}{2\pi^2}\,\gamma_i\, g_i\, m_i^2 \, T(\tau)\, 
	K_2\!\left( \frac{m_i}{T(\tau)}\right),
	\label{statistical}
\end{align}
In this expression, $\gamma_i$, $g_i$, and $m_i$ denote the fugacity, degeneracy, and mass of particle $i$, respectively, while $T(\tau)$ is the time-dependent temperature. The abundance $N_i(\tau)$ is obtained as the product of the number density $n_i(\tau)$ and the volume $V(\tau)$. The fugacities are adjusted so as to reproduce the multiplicities listed in Table~\ref{tabela1}.
For charm quarks, however, which are produced in the early stages of the collision, the total number $N_c$ in charmed hadrons is assumed to remain constant throughout the hadronic gas phase. This requirement leads to a time-dependent charm fugacity factor $\gamma_c \equiv \gamma_c(\tau)$, introduced to preserve $N_c = n_c(\tau) \times V(\tau) = 14$. Consequently, the abundances of the $D$, $\bar D$, $D^*$, and $\bar D^*$ mesons incorporate the resulting factor $\gamma_c(\tau)$.

To model the hadron gas evolution, we adopt the boost-invariant Bjorken scenario with accelerated transverse expansion~\cite{Bjorken:1982qr}. Hydrodynamical calculations indicate that both the temperature and the volume evolve with time during the hadronic phase of HICs. These quantities can be parametrized as functions of the proper time $\tau$ in the following way~\cite{cho2013hadronic,XProd1,ABREU2016303,Abreu:2017cof,Abreu:2018mnc,magalhaes_2024}:
\begin{align}
	V(\tau) & = \pi \left[ R_C + v_C \left(\tau - \tau_C\right) + 
	\frac{a_C}{2} \left(\tau - \tau_C\right)^2 \right]^2 \tau c , \nonumber \\
	T(\tau) & =  T_C - \left( T_H - T_F \right) \left( \frac{\tau - 
		\tau _H }{\tau _F - \tau _H}\right)^{\alpha}.
	\label{eq: TempVol}
\end{align}
Here, \( R_C \), \( \upsilon_C \), \( a_C \), and \( T_C \) denote, respectively, the transverse size, transverse velocity, transverse acceleration, and temperature at the critical time \( \tau_C \); \( T_H \) is the hadronization temperature at time \( \tau_H \), and \( T_F \) is the kinetic freeze-out temperature at time \( \tau_F \). These parameters, which are fixed according to Ref.~\cite{Cho:2017dcy} for a hadronic medium formed in central Pb--Pb collisions at \( \sqrt{s_{NN}} = 5.02 \,\mathrm{TeV} \), are shown in Table~\ref{tabela1}.


\begin{table}[htb!]
	\caption{The parameters employed in Eq.~(\ref{eq: TempVol}) for central Pb--Pb collisions at $\sqrt{s_{NN}} = 5.02 \,\mathrm{TeV}$ extracted from Ref.~\cite{Cho:2017dcy}, along with the initial multiplicities of the mesons and charm quarks, and the coalescence-model frequency $\omega_G$.	}
	\centering \begin{tabular}{ccc}
		\hline \hline $v_C(\mathrm{c})$ & $a_C\left(\mathrm{c}^2 / \mathrm{fm}\right)$ & $R_C(\mathrm{fm})$ \\
		0.5 & 0.09 & 11 \\
		\hline $\tau_C(\mathrm{fm} / \mathrm{c}) $&$ \tau_H(\mathrm{fm} / \mathrm{c}) $&$ \tau_F(\mathrm{fm} / \mathrm{c})$ \\
		7.1 & 10.2 & 21.5 \\
		\hline $T_C(\mathrm{MeV}) $&$ T_H(\mathrm{MeV}) $&$ T_F(\mathrm{MeV})$ \\
		156 & 156 & 115 \\
		\hline $N_\pi\left(\tau_H\right) $&$\omega_G[\mathrm{MeV}]  $& $\alpha$\\
		713 &19.8  &  $0.8$ \\
		\hline $N_D\left(\tau_H\right) $&$N_{D^*}\left(\tau_H\right)  $& $N_c$\\
		4.7 & 6.3  &  $14$ \\
		\hline \hline \label{tabela1}
	\end{tabular}
\end{table}


To determine the initial condition for the rate equation~(\ref{rateequation}) of the $G(3900)$ multiplicity, we employ two approaches: the statistical hadronization model (SHM) and the coalescence model (COM).

In the first case, we make use of Eq.~\eqref{statistical}, taking into account that the quark content of the $G(3900)$ includes a charm--anti-charm quark pair, yielding the factor $\gamma_G(\tau) \equiv \gamma_c(\tau) ^{n_c + n_{\bar c}} = \gamma_c ^2(\tau)$, which gives us
\begin{equation}
	N_{G}^{SHM}=N_G(\tau_H) \approx 2.3\times 10^{-3}.
    \label{initialconditionsSHM}
\end{equation}



In the case of the COM, the multiplicity of the hadronic state is obtained by convolving the density matrix of the constituents with their Wigner function, which retains essential information about the system's intrinsic structure---such as angular momentum, and the type and number of constituent quarks. As discussed previously, and assuming the $G$ to be a $P$-wave $D\bar{D}^{*}/\bar{D}D^{*}$ molecular state, the yield of the $G(3900)$ at $\tau_c$ can be expressed as~\cite{EXHIC,Cho:2017dcy}:
\begin{equation}
	N_G^{COM}  \approx g_G\frac{N_D}{g_D}\frac{N_{D^*}}{g_{D^*}}\frac{(4 \pi \sigma_i^2)^{\frac{3}{2}}}{V \bigl(1 + 2 \mu_i T \sigma_i^2 \bigr)}\times\left[ \frac{4 \mu_i T \sigma_i^2}{3 \bigl(1 + 2 \mu_i T \sigma_i^2 \bigr)}
	\right],
    \label{GCOM}
\end{equation}
where $g_j$ and $N_j$ are, respectively, the degeneracy and the number of the $j$-th constituent of $G$, $g_G$ is the degeneracy of $G$, and $\sigma_i = (\mu_i \omega_G)^{-1/2}$. The internal structure of the hadron is assumed to be represented by a harmonic oscillator, where $\omega_G$ is the oscillator frequency and $\mu_i$ is the reduced mass. The value of $\omega_G$ used in this work, shown in Table~\ref{tabela1}, was determined via $\omega_G = 6 E_b \bigl[m_{D^*} + m_D - m_G\bigr]$, with $E_b = m_{D^*} + m_D - m_G$ the binding energy of the $G(3900)$. Substituting these parameters into Eq.~(\ref{GCOM}) together with the constants listed in Table~\ref{tabela1}, the initial multiplicity of $G$ in accordance with the COM is found to be
\begin{align}
	N_{G}^{COM}=N_{G}(\tau_H) \approx  6.6\times 10^{-3}.
	\label{initialconditionsCoal}
\end{align}

In the next section we will present the obtained results.

\subsection{Results and discussion}

\begin{figure}[!htbp]
\centering
\includegraphics[{width=1\linewidth}]{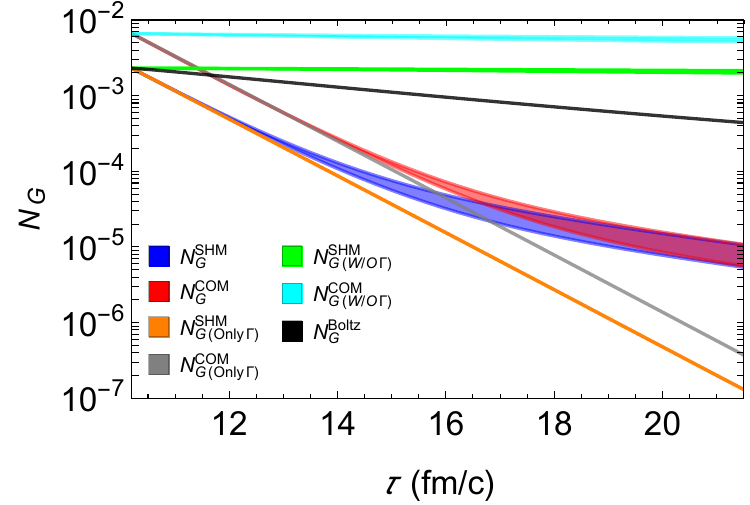}
\caption{ Abundance of $G (3900)$ as a function of the proper time in central Pb-Pb collisions at $\sqrt{s_{NN}} = 5.02$ TeV. The curves correspond to initial conditions calculated from the statistical and coalescence models, with all contributions in the right hand side of Eq.~\eqref{rateequation} (denoted as $N_{G}$); without the decay width contribution in the last line of the expression ($N_{G(\mathrm{W/O} \ \Gamma)}$); and considering only  decay width contribution ($N_{G(\mathrm{Only} \ \Gamma)}$). The Boltzmann distribution obeying Eq.~\eqref{statistical} is also included ($N_{G}^{Boltz})$). }
    \label{TimeEvNZ}
\end{figure}

In Fig.~\ref{TimeEvNZ}, we show the time evolution of the $G(3900)$ multiplicity obtained by solving Eq.~\eqref{rateequation}. For comparison, several scenarios are explored: curves obtained using initial conditions from the SHM (Eq.~\eqref{initialconditionsSHM}) and the coalescence model (COM, Eq.~\ref{initialconditionsCoal}) while considering all terms on the right-hand side of Eq.~\eqref{rateequation}; curves obtained without the decay contribution (last term in the expression); and curves considering only the decay width contribution. The Boltzmann distribution obeying Eq.~\eqref{statistical} is also included for reference.

In both the SHM and COM scenarios, omitting the decay width contribution leads to no significant change in the $G$ multiplicity. In this scenario, the hadronic phase does not substantially alter the final abundance of $G$, with the system retaining memory of its initial multiplicity. Furthermore, the final multiplicity is of the same order of magnitude ($\sim 10^{-3}$) in both models, with the SHM value being approximately three times smaller than that of the COM.

When the decay contribution is taken into account, however, a distinct behavior emerges: the $G$ multiplicity is strongly suppressed by approximately three orders of magnitude. This suppression arises from the large width of the state, which renders its lifetime sufficiently short for decay to occur during the fireball's lifetime. This behavior also differs from that of the Boltzmann distribution. Interestingly, during the mid-to-late hadron gas phase, the curves for the SHM and COM scenarios deviate from the pure decay-width-only case. They decay more slowly and converge to similar final values. This behavior indicates that the gain terms in Eq.~\eqref{rateequation} become increasingly relevant as the hadron gas phase evolves.

It is instructive to compare these findings with other results reported in the literature for this and similar exotic states. In Ref.~\cite{Cao:2025okk} 
the authors employed a parton and hadron cascade model to simulate the production of $G(3900)$ in $e^+e^-$ annihilations at $\sqrt{s}=4.95$ GeV. Identifying the structure as either a $P$-wave $D\bar{D}$ or $D\bar{D}^*/\bar{D}D^*$ state, they  obtained multiplicities of $N_G = 3.87 \times 10^{-4}$ and $6.08 \times 10^{-4}$, respectively. While our initial conditions (Eqs.~\eqref{initialconditionsSHM} and \eqref{initialconditionsCoal}) give higher multiplicities than those estimated in Ref.~\cite{Cao:2025okk}, our final multiplicities are smaller. This indicates that, under the hadronic molecule interpretation, detecting the $G$ in heavy-ion collisions is challenging, and its production is easier in other collision environments.

In addition, we estimate the ratio of the $G(3900)$ multiplicity to that of the $Z_c(3900)$. It should be stressed that the probability of the $S$-wave $D\bar{D}^*$ component of the $Z_c(3900)$, calculated via the compositeness condition, was found to be less than $0.5$. This suggests that other hadronic components, compact quark-state cores, or kinematic effects may also play an important role in the structure of the $Z_c(3900)$ (see discussion in Ref.~\cite{magalhaes_2024}). Using the final yield of the $Z_c(3900)$ in central Pb-Pb collisions at $\sqrt{s_{NN}} = 5.02$ TeV from Ref.~\cite{magalhaes_2024}, where the $Z_c(3900)$ was considered as a compact tetraquark state, we obtain
\begin{align}
\frac{N_G^{\text{(COM)}}}{N_{Z_c}} & \approx 4\text{--}8 \times 10^{-3}, \nonumber \\
\frac{N_G^{\text{(SHM)}}}{N_{Z_c}} & \approx 3\text{--}7 \times 10^{-3}.
\label{ratioGZc}
\end{align}
Thus, we see that the multiplicity of $G$ is suppressed relative to that of $Z_c$, under both the SHM and COM scenarios.
The above numbers can be considered upper bounds. Although we do not have the abundancies of $Z_c(3900)$ treated as a meson molecule, we 
know from previous works that the abundances of molecules are always larger than those of tetraquarks. 
The $G$ suppression can be attributed 
to its large decay width. A larger decay width implies a shorter lifetime, which increases the probability that the $G(3900)$ decays during the fireball lifetime relative to the $Z_c(3900)$.




\section{Concluding remarks}
\label{Conclusions}

In this work, we investigated the interactions of the exotic hadron $G(3900)$---interpreted as a $P$-wave $D\bar{D}^*/D^*\bar{D}$ molecular state---with a hadronic medium composed of light mesons produced in heavy-ion collisions (HICs). Using an effective Lagrangian approach to estimate vacuum cross sections and their thermal averages, we found that absorption processes dominate over production processes for the $G$ state.

We then studied the time evolution of the $G$ multiplicity under two different initial conditions derived from the statistical hadronization model (SHM) and the coalescence model (COM). Our results indicate that the $G(3900)$ is significantly affected by the hadronic medium formed in HICs, as evidenced by the strong suppression of its abundance as a function of proper time $\tau$. The final yield of $G$ loses memory of its initial production mechanism, mainly due to its large width, which renders its lifetime sufficiently short for decay during the hadron gas phase.

The comparison of our findings with related studies has also been done. Ref.~\cite{Abreu:2024mxc} shows that when the $Z_c$ is considered a tetraquark state, the combined effects of hadronic interactions, hydrodynamical expansion, decay, and regeneration lead to an enhancement of its multiplicity by the end of the collision. Consequently, we obtain a considerably smaller final yield of $G$ compared to that of the $Z_c(3900)$.

To place our results in a broader perspective, we note that in Refs.~\cite{Abreu:2020jsl,Llanes-Estrada:2021ath} it was argued that cusps arising from kinematic effects---such as triangle singularities---are erased in the HIC environment when their widths are sufficiently large, as is the case for the $Z_c$. Accordingly, detection of the $Z_c$ in HICs would favor its existence as a genuine hadron, while its absence would support a kinematic origin. In this context, we recall that an alternative mechanism for generating the $G$ via a $D_1 D^* D^*$ triangle loop~\cite{Huang:2025g3900} was discussed in the Introduction. Given the disappearance of triangle singularities in HICs for broad states~\cite{Abreu:2024mxc}, it follows that the detection of $G$ remains disfavored in \textit{both} the hadronic molecule and the triangle singularity scenarios. This constitutes a clear distinction from the case of the $Z_c(3900)$.

\begin{acknowledgements}
This work was partly supported  by the Brazilian agencies CNPq/FAPERJ under the Project INCT-Física Nuclear e Aplicações (Contract No. 408419/2024-5). The work of L.M.A. is partly supported by the Brazilian agency CNPq (Grants No. 400215/2022-5, 308299/2023-0, 402942/2024-8).    


\end{acknowledgements} 


\appendix

\section{Calculation of the coupling $g_{G_0DD^{*}}$}
\label{appa}

We interpret the $G(3900)$ with $I^G(J^{PC}) = 0^-(1^{--})$ as a $P$-wave $D\bar{D}^*/D^*\bar{D}$ bound state. Therefore, in consonance with Refs.~\cite{Entem:2011nv,Ortega:2012rs,Du:2016qcr,Lin:2024qcq, Chen:2025gxe, Ye:2025ywy, Chen:2025mgg}, its structure is dominated by the $D\bar{D}^{\ast}/\bar{D}D^{\ast}$ channels and expressed as:
\begin{equation}
|G\rangle = \frac{1}{\sqrt{2}} \bigl( |D\bar{D}^*\rangle +|\bar{D}D^*\rangle \bigr),
\label{G_0}
\end{equation}
with
\begin{equation}
|D\bar{D}^*\rangle = \frac{1}{\sqrt{2}} \bigl( |D^+\bar{D}^{*-}\rangle - |D^0\bar{D}^{*0}\rangle \bigr).
\label{G_0bis}
\end{equation}

\begin{figure}[h]
    \centering
\includegraphics[width=0.4\textwidth]{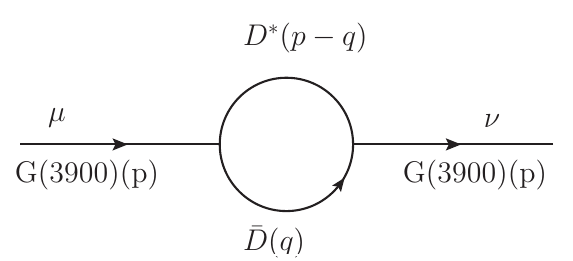}
\caption{Feynman diagram contributing to the self-energy $\Sigma^{\mu\nu}(p)$ of the $ G $ state. }
    \label{autoenergy}
\end{figure}

In this scenario, we determine the coupling $g_{GDD^*}$ in the effective Lagrangian~(\ref{lagrangianas})  via Weinberg's compositeness criterion~\cite{Weinberg:1965zz}. Accordingly, we use the non-local version of $\mathcal{L}_{GDD^*}$ as follows:
\begin{align}
\mathcal{L}_{GDD^*}^{\rm NL}(x)
&=
-g_{GDD^*}\,
\epsilon^{\mu\nu\rho\sigma}
\,\partial_\mu G_\nu(x)
\int d^4y\,\Phi(y^2)\nonumber \\
&\times\Big[
D\!\left(x+w_{D^*}y\right)
\overset{\leftrightarrow}{\partial_\rho}
\bar{D}^{*}_{\sigma}\!\left(x-w_{D}y\right)\nonumber \\
&+
D^{*}_{\sigma}\!\left(x+w_{D}y\right)
\overset{\leftrightarrow}{\partial_\rho}
\bar{D}\!\left(x-w_{D^*}y\right)
\Big],
\label{NLLAG}
\end{align}
where $x$ and $y$ correspond to the center-of-mass and the relative Jacobi coordinates of the internal molecular constituents $D D^*$, respectively~\cite{Dong:2017gaw,Liu:2025sjz,Zhu:2021exs}. The kinematic weights are defined as $\omega_{D(D^*)} = m_{D(D^*)}/(m_{D(D^*)} + m_{D^*(D)})$, with $m_{D(D^*)}$ the mass of the meson $D(D^*)$. In addition, $\Phi(y^2)$ is the correlation function that encodes the finite spatial extension of the bound state. Its corresponding momentum-space representation is obtained through the Fourier transform
\begin{equation}
\Phi(y^2) = \int \frac{d^4q}{(2\pi)^4} e^{-iqy} , \tilde{\Phi}(-q^2),
\label{eq:fft}
\end{equation}    
In order to control ultraviolet divergences, we choose the gaussian form: 
\begin{equation}
\tilde{\Phi}(q_E^2) = e^{-q_E^2 / \Lambda_g^2},
\label{eq:ff}
\end{equation}
where $q_E$ is the Euclidean four-momentum and $\Lambda_g$ is the size parameter (cutoff). According to the Weinberg compositeness condition~\cite{Weinberg:1965zz}, the coupling constant $g_{GDD^*}$ is determined by setting the wavefunction renormalization constant of the composite state $G$ to zero:
\begin{eqnarray}
    Z = 1 - \frac{d\Sigma}{dp^2_{G}}\bigg|_{p^2 = m^2_{G}} = 0,
\label{WBcomp1}
\end{eqnarray}
where $\Sigma$ is the transverse part of the mass operator (self-energy) $\Sigma^{\mu\nu}$ extracted from Eq.~(\ref{NLLAG}), which is represented diagrammatically in Fig.~\ref{autoenergy}.

After some algebra involving the totally antisymmetric Levi-Civita tensor, the four-momenta, and the metric tensor, we find
\begin{align}
    i\Sigma^{\mu\nu}(p) & = i^2g_{GDD^*}^2\int\frac{d^4q}{(2\pi)^4}\frac{N^{\mu\nu}(q)}{[q^2-m_{\bar{D}}^2][(p-q)^2-m_{D^*}^2]}\nonumber \\
    &\times\Tilde{\Phi}^2([q - \omega_{\bar{D}} p]^2).
\label{mass_op_0}
\end{align}
Its transverse part is obtained from the standard decomposition
\begin{align}
    \Sigma^{\mu\nu}(p) &= g_{\perp}^{\mu\nu}\Sigma(p^2) + \frac{p^\mu p^{\nu}}{p^2}\Sigma^{L}(p^2) \nonumber \\
    &= \bigg(g^{\mu\nu}-\frac{p^\mu p^\nu}{p^2}\bigg)\Sigma(p^2) + \frac{p^\mu p^{\nu}}{p^2}\Sigma^{L}(p^2),
    \label{massop}
\end{align}
where $\Sigma^{L}$ is the longitudinal part. The tensor $N^{\mu\nu}(q)$ is given by:
\begin{align}
    N^{\mu\nu}(q) = &-\bigg[p^2q^\mu q^\nu-(p\cdot q)(p^\nu q^\mu + p^\mu q^\nu)+(p\cdot q)^2g^{\mu\nu} \nonumber \\
    &+q^2(p^\mu p^\nu - p^2g^{\mu\nu})\bigg]. 
\end{align}
We employ the Schwinger parametrization for the propagators in Eq.~(\ref{mass_op_0}):
\begin{align}
    \frac{1}{q^2-m_{\bar{D}}^2} &= \int_0^\infty d\alpha\exp\left\{\alpha[q^2-m_{\bar{D}}^2]\right\} ,  \nonumber \\
    \frac{1}{(p-q)^2-m_{D^*}^2} &= \int_0^\infty d\beta\exp\left\{\beta[(p-q)^2-m_{D^*}^2]\right\},  
\label{Schwingerpar}
\end{align}
where the parameters $\alpha$ and $\beta$ have dimension of  $[\text{energy}]^{-2}$. It is convenient to perform the rescaling $\alpha\to\alpha/\Lambda_g^2$ and $\beta\to\beta/\Lambda_g^2$. After that, Eqs.~(\ref{eq:fft}) and~(\ref{Schwingerpar}) allows us to write the mass operator as:
\begin{align}
    i\Sigma^{\mu\nu}(p) & = -\frac{g_{GDD^*}^2}{\Lambda_g^4}\int\frac{d^4q}{(2\pi)^4}\int_{0}^{\infty}d\alpha d\beta\exp[f(q,\alpha,\beta)]\nonumber \\
    &\times N^{\mu\nu}(q),
\label{mass_op_2}
\end{align}
where 
\begin{align}
f(q,\alpha,\beta) &= \frac{1}{\Lambda_g^2}[\alpha q^2 - \alpha m_{\bar{D}}^2 + \beta p^2 -2\beta(p\cdot q) - \beta m_{D^*}^2 \nonumber \\
&+ \beta q^2 + 2q^2 -4\omega_{\bar{D}}(p\cdot q) + 2\omega_{\bar{D}}^2p^2],\nonumber \\
&= \frac{1}{\Lambda_g^2}[q^2z - 2(p\cdot q)(\beta + 2\omega_{\bar{D}})+ p^2(\beta +  2\omega_{\bar{D}}^2) \nonumber \\
&- \alpha m_{\bar{D}}^2 - \beta m_{D^*}^2], \nonumber
\end{align}
with $z(\alpha,\beta) = (\alpha + \beta + 2)$. In order to have a more tractable expression, we rewrite the function $f(q,\alpha,\beta)$ by completing the square in the variable $q$:
\begin{align*}
    f(q,\alpha,\beta) &= \frac{1}{\Lambda_g^2}\bigg\{z(\alpha,\beta)\bigg[q- \frac{2p(\beta + 2\omega_{\bar{D}})}{2z(\alpha,\beta)}\bigg]^2 \nonumber \\
    &-\frac{[2p(\beta + 2\omega_{\bar{D}})]^2}{4z(\alpha,\beta)}+p^2(\beta +  2\omega_{\bar{D}}^2) - \alpha m_{\bar{D}}^2 - \beta m_{D^*}^2\bigg\},  
\end{align*}
and making the change of variable
\begin{align}
    u^2 &= \frac{z}{\Lambda_g^2}\bigg[q+ \frac{p\Delta}{2z(\alpha,\beta)}\bigg]^2,
\label{transfqtou}
\end{align}
where for a clearer notation we defined $\Delta = -2(\beta + 2\omega_{\bar{D}})$. 
Therefore, the mass operator assumes the form:
\begin{align}
        i\Sigma^{\mu\nu}(p) &= -g_{GDD^*}^2\int_{0}^{\infty}d\alpha d\beta\frac{\exp[g(p,\alpha,\beta)]}{{z(\alpha,\beta)^2}}\nonumber \\
        &\times\int\frac{d^4u}{(2\pi)^4}\exp[u^2]\times N^{\mu\nu}(u), 
\label{massop2}
\end{align}
where
\begin{align}
   g(p,\alpha,\beta) = -(1/\Lambda_g^2)[\Delta^2/4z(\alpha,\beta)&-p^2(\beta +  2\omega_{\bar{D}}^2) \nonumber \\
    &+ \alpha m_{\bar{D}}^2 + \beta m_{D^*}^2 ].\nonumber \\
\end{align}
The next step is to perform a Wick rotation in order to work with the Euclidean variable $u_E$, allowing us to put the second line of Eq.~(\ref{massop2}) in the standard Gaussian integration form. In this sense, we set $u^2\to-u_E^2$ and $d^4u\to id^4u_E$. In addition, we just pick the transverse part of the mass operator, which means that we should split the tensor $N^{\mu\nu}$ in the form of Eq.~(\ref{massop}). As a consequence, the tensor structure becomes considerably simplified. We thus have  
\begin{align}
   i\int\frac{d^4u_{E}}{(2\pi)^4}\exp[-u_{E}^2]\times\bigg[\frac{-4p^2\Lambda_g^2u_E^2}{z(\alpha,\beta)}\bigg] = -\frac{(2\pi^2)}{(2\pi)^4}\frac{4p^2\Lambda_g^2}{z(\alpha,\beta)}i.
\end{align}
Finally,
\begin{align}
           \Sigma(p^2)  &=\frac{g_{GDD^*}^2}{2\pi^2}\int_{0}^{\infty}d\alpha d\beta\exp\{-(1/\Lambda_g^2)[\Delta^2/4z(\alpha,\beta) \nonumber \\
           &-p^2(\beta +  2\omega_{\bar{D}}^2) + \alpha m_{\bar{D}}^2 + \beta m_{D^*}^2 ]\}\bigg(\frac{p^2\Lambda_g^2}{z(\alpha,\beta)^3}\bigg).
\end{align}
After numerical integration, we use Eq.~(\ref{WBcomp1}) and find: 
\begin{align}
g_{GDD^*}^{(1\,\mathrm{GeV})} &= 4.01~\text{GeV$^{-1}$} \nonumber \\ g_{GDD^*}^{(0.6\,\mathrm{GeV})} &= 5.71~\text{GeV$^{-1}$} 
\end{align}
These results were used to estimate the uncertainty band in our results. 
We should also note that the relations between this coupling in the particle basis to those in the isospin basis according to Eqs.~\eqref{G_0} and~\eqref{G_0bis}.


\bibliographystyle{apsrev4-2}
\bibliography{references-m-G3900}

\end{document}